\documentclass[runningheads]{llncs}
\usepackage[T1]{fontenc}
\usepackage{graphicx,verbatim}
\usepackage{colortbl}
\usepackage{xcolor} 
\usepackage{multirow}
\usepackage{makecell}
\usepackage{siunitx}
\usepackage{xspace}
\usepackage[normalem]{ulem}
\usepackage{paralist}
\usepackage{hyperref}

\makeatletter
\DeclareRobustCommand\onedot{\futurelet\@let@token\@onedot}
\def\@onedot{\ifx\@let@token.\else.\null\fi\xspace}
\def\eg{\emph{e.g}\onedot} 
\def\ie{\emph{i.e}\onedot}

\def\etal{\emph{et al}\onedot}

\makeatother

\begin{document}
%

\title{A Novel Framework for Integrating \\3D Ultrasound into Percutaneous Liver Tumour Ablation}

%
\author{Shuwei Xing\inst{1,2}$^{*}$
\and
Derek W. Cool\inst{5}
\and
David Tessier\inst{1}
\and
Elvis C.S. Chen\inst{1,2,3,4}
Terry M. Peters\inst{1,2,3}
\and
Aaron Fenster\inst{1,2,3}
}
\authorrunning{S. Xing et al.}
%



\institute{Robarts Research Institute, Western University, London, Canada 
\and School of Biomedical Engineering, Western University, London, Canada
\and Department of Medical Biophysics, Western University, London, Canada 
\and Department of Electrical and Computer Engineering, Western University, London, Canada
\and Department of Medical Imaging, Western University, London, Canada
\\
    \email{xshuwei@uwo.ca}}

\maketitle              
\begin{abstract}
3D ultrasound (US) imaging has shown significant benefits in enhancing the outcomes of percutaneous liver tumour ablation. Its clinical integration is crucial for transitioning 3D US into the therapeutic domain. However, challenges of tumour identification in US images continue to hinder its broader adoption. In this work, we propose a novel framework for integrating 3D US into the standard ablation workflow. We present a key component, a clinically viable 2D US–CT/MRI registration approach, leveraging 3D US as an intermediary to reduce registration complexity. To facilitate efficient verification of the registration workflow, we also propose an intuitive multimodal image visualization technique. In our study, 2D US–CT/MRI registration achieved a landmark distance error of $\sim$2–4 \si{\milli\meter} with a runtime of \qty{0.22}{\second} per image pair. Additionally, non-rigid registration 
reduced the mean alignment error by $\sim$40\% compared to rigid registration. Results demonstrated the \textcolor{black}{efficacy} of the proposed 2D US–CT/MRI registration workflow. Our integration framework advanced the capabilities of 3D US imaging in improving percutaneous tumour ablation, demonstrating the potential to expand the therapeutic role of 3D US in clinical interventions.


\end{abstract}
\section{Introduction}
Percutaneous tumour ablation, including radiofrequency and microwave ablation, is an effective minimally invasive treatment for early-stage liver cancer~\cite{lencioni2005early,sala2004initial}. This technique involves inserting one or more \textcolor{black}{ablation} applicators into the tumour's centroid, generating a thermal ablation zone around the applicator \textcolor{black}{active} tip to induce irreversible cellular death. Clinically, ultrasound (US) is widely used for applicator guidance due to its real-time imaging capabilities and broad accessibility. However, achieving accurate applicator placement requires physicians to mentally reconstruct 3D anatomical structures, such as tumours and surrounding vessels, from \textcolor{black}{series of} 2D US images. This cognitive burden, combined with the reliance on extensive experience and training, presents challenges in ensuring consistent outcomes in US-guided ablation. 3D US imaging has gained considerable attention for enhancing tumour ablation. Compared to conventional 2D US, it reduces operator dependence and enables volumetric measurement~\cite{fenster2001three}. Recent work has demonstrated that 3D US imaging provides better tumour coverage than 2D US alone, as evidenced by corresponding patient outcomes~\cite{xing20223d}. For effective integration into the standard workflow, 3D US is recommended for assessing intra-procedural tumour coverage, while 2D US remains the primary real-time applicator guidance modality. Nevertheless, challenges in identifying tumours in US images continue to hinder the widespread adoption of this workflow. \textcolor{black}{Specifically}, \textcolor{black}{certain types of} liver tumours have low conspicuity or are nearly undetectable when imaging small lesions or those in difficult-to-image regions (\eg, the liver dome). Moreover, tumour mimics, such as regenerative nodules in cirrhotic livers and prior ablation sites~\cite{elsayes2018spectrum}, complicate tumour identification and procedural accuracy.

To address these limitations, image registration approaches that align 2D US with CT or MRI, have been investigated~\cite{fuerst2014automatic,wein2008automatic,pardasani2016single}. For example, the ``Linear Correlation of Linear Combination'' (LC2) similarity metric was developed to facilitate automatic US–CT~\cite{wein2008automatic} and US–MRI~\cite{fuerst2014automatic} registration. 
Extrinsic spatial trackers, including optical~\cite{penney2004registration} and magnetic~\cite{paccini2024novel}, have been used to assist US–CT/MRI alignment, although they often fail to account for the respiratory motion during the intervention. While these approaches demonstrated significant technological advancements, they have not yet been standardised for integration into the ablation workflow due to their complexity and computational demands. 

Recently, deep learning techniques have been \textcolor{black}{developed} to \textcolor{black}{facilitate these} registration challenges. Markova \etal~\cite{markova2022global} proposed a computationally efficient end-to-end registration approach that aligns extracted keypoints from US and MRI liver images using an adapted LoFTR~\cite{sun2021loftr} network. Azampour \etal~\cite{azampour2023self} developed a deep regression model for US probe pose estimation. In this framework, a US synthesis module was integrated to address the domain gap between US and CT images. However, \textcolor{black}{existing} learning approaches \textcolor{black}{are unable} to achieve clinically acceptable registration accuracy and remain far from clinical deployment. Inspired by a strategy employed in prostate interventions~\cite{natarajan2011clinical,xu2008real,guo2022ultrasound}, 2D US–CT/MRI registration can be divided into two sequential registration steps: ``3D US–CT/MRI'' and ``2D–3D US'', reducing alignment complexity. For example, Xu \etal~\cite{xu2008real} demonstrated that transrectal US images can be fused with pre-procedural MRI for prostate biopsy using 3D US as an intermediary to align 2D transrectal US with MRI. This registration decomposition shows promise for integration into the standard workflow; however, its applicability to liver interventions has yet to be explored in depth.

This paper presents a 3D US system integrated with registration and visualization for percutaneous liver tumour ablation. We advances 3D US imaging beyond its diagnostic role to therapeutic applications in liver cancer. Our contributions include \begin{inparaitem} 
 \item[1)] an effective 3D US integration framework compatible with the standard ablation workflow,
 \item[2)] a clinically viable 2D US–CT/MRI registration workflow optimized for accuracy, runtime and robustness, and
 \item[3)] an intuitive multimodal image visualization tool for validating intra-procedural alignment.
\end{inparaitem}

\section{Methods: 3D US Integration Framework}

\begin{figure}[htbp]
\centering
\includegraphics[width=0.95\textwidth]{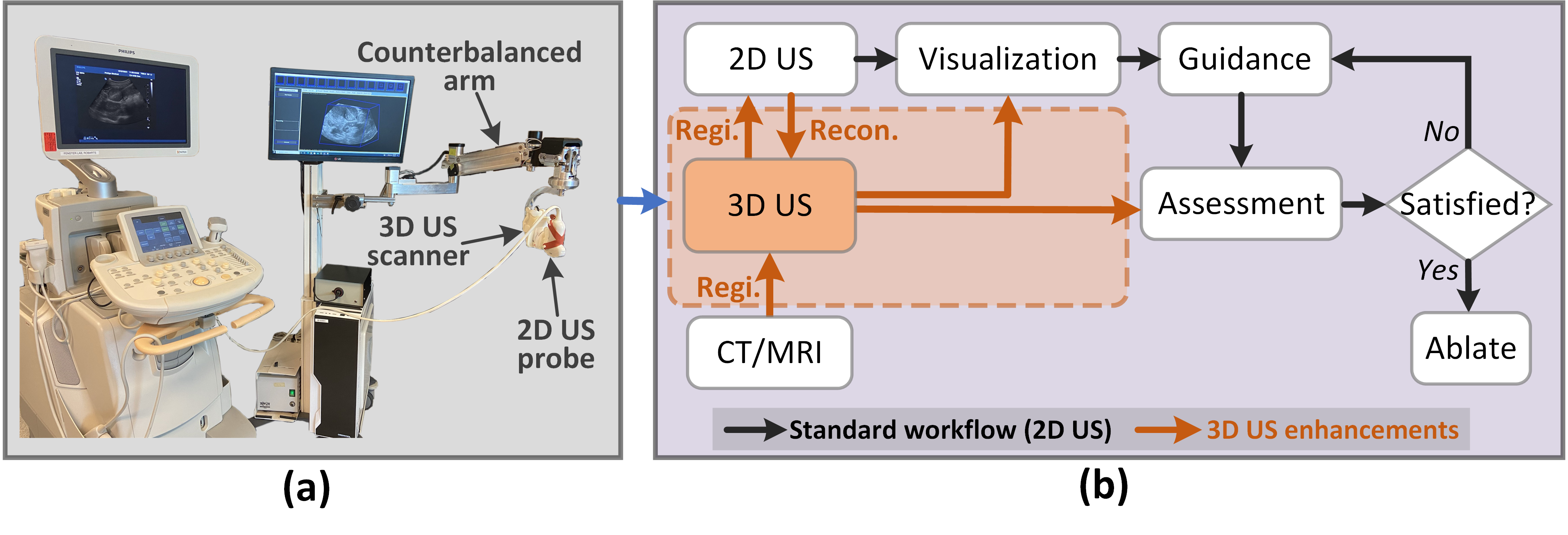}
\caption{Framework for 3D US integration into 2D US guidance. (a) 3D US system and (b) intra-procedural guidance. The role of 3D US is highlighted in a brown block. (Regi.: Registration, Recon.: 3D US reconstruction)} \label{fig_workflow}
\end{figure}

\noindent Fig.~\ref{fig_workflow}a depicts the proposed mechatronic 3D US system, including a counterbalanced arm equipped with joint encoders, and a motor-driven 3D US scanner at its end. This device can be integrated with any commercial 2D US machine compatible with a liver-imaging US probe. The 3D US scanner enables a 2D US probe to perform automated movements, including tilting (\qty{60}{\degree}), translation (\qty{60}{\milli\meter}), and hybrid rotation-translation motion. The acquired US frames are then used to reconstruct a 3D US image within \qtyrange{7}{12}{\second}. The integration framework of our 3D US system consists of a pre-procedural spatial calibration (Sec.~\ref{sec_calibration}) and an optimized intra-procedural US guidance workflow (Fig.~\ref{fig_workflow}b). The standard ablation workflow is enhanced with US–CT/MRI registration, multimodal visualization, and tumour coverage assessment, as highlighted in the brown block. Given our previous work on tumour coverage assessment~\cite{xing20223d}, this work primarily focuses on 2D US–CT/MRI registration and visualization (Sec.~\ref{sec_registration}).

\begin{figure}[htbp]
\centering
\includegraphics[width=0.85\textwidth]{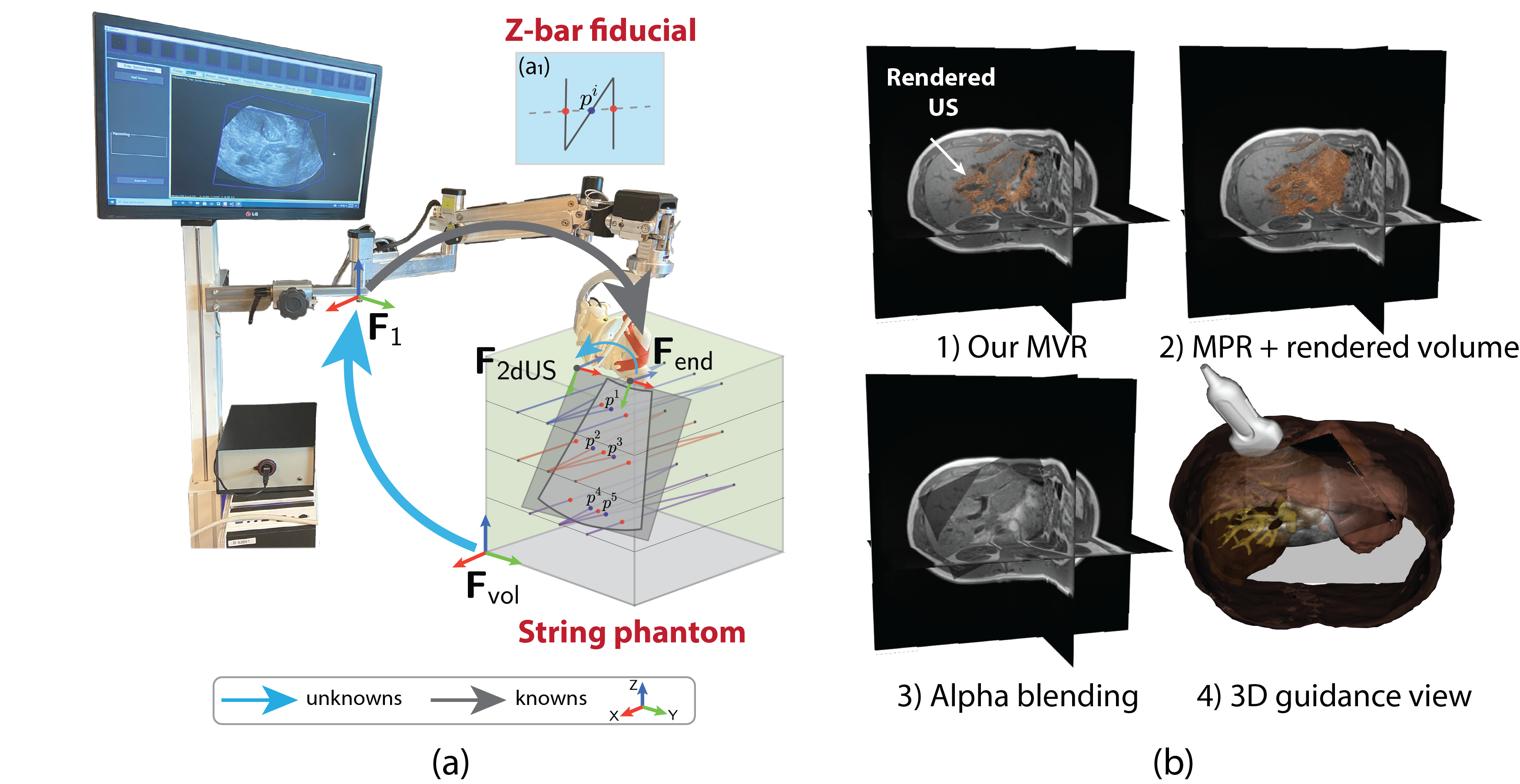}
\caption{System calibration and visualization. (a) US probe calibration, and (b) different visualization approaches. (MVR: Multi-planar and Volume Rendering; MPR: Multi-planar Reformation)} \label{fig_calibration_visualization}
\end{figure}

\subsection{System Calibration}\label{sec_calibration}


\subsubsection{Mechatronic Arm Calibration}
This calibration determines the pose of the physical end of the US probe ($F_{end}$) relative to the base of the arm ($F_1$) (Fig.~\ref{fig_calibration_visualization}a). Following Li \etal~\cite{li2019kinematics}, we first modelled the kinematics \textcolor{black}{of the mechatronic arm} using the Denavit-Hartenberg approach~\cite{denavit1955kinematic}, then employed a Levenberg-Marquardt nonlinear regression model~\cite{marquardt1963algorithm} to compute the calibration parameters. Notably, to enable tracking of the probe end using the optical tracker (Polaris Spectra, Northern Digital Inc., Canada), we designed a US probe-mimicking phantom that incorporated a dynamic reference body.

\subsubsection{US Probe Calibration}
This step establishes the spatial relationship between $F_{end}$ and the 2D US image frame ($F_{2dUS}$).  For this purpose, we first designed a string phantom encased in agar to simulate background tissue. The phantom contained three layers of nylon wires, arranged in 15 Z-bar fiducials~\cite{comeau1998integrated} (Fig.~\ref{fig_calibration_visualization}a). Using triangle similarity, the middle point ($p^i$), where the Z-bar fiducial intersects the US image plane, can then be used to derive its corresponding 3D position relative to the phantom frame ($F_{vol}$). Finally, we applied hand-eye calibration~\cite{tsai1989new} to compute the transformation from $F_{end}$ to $F_{2dUS}$.

\subsection{2D US–CT/MRI Registration}\label{sec_registration}

\begin{figure}[h]
\centering
\includegraphics[width=0.85\textwidth]{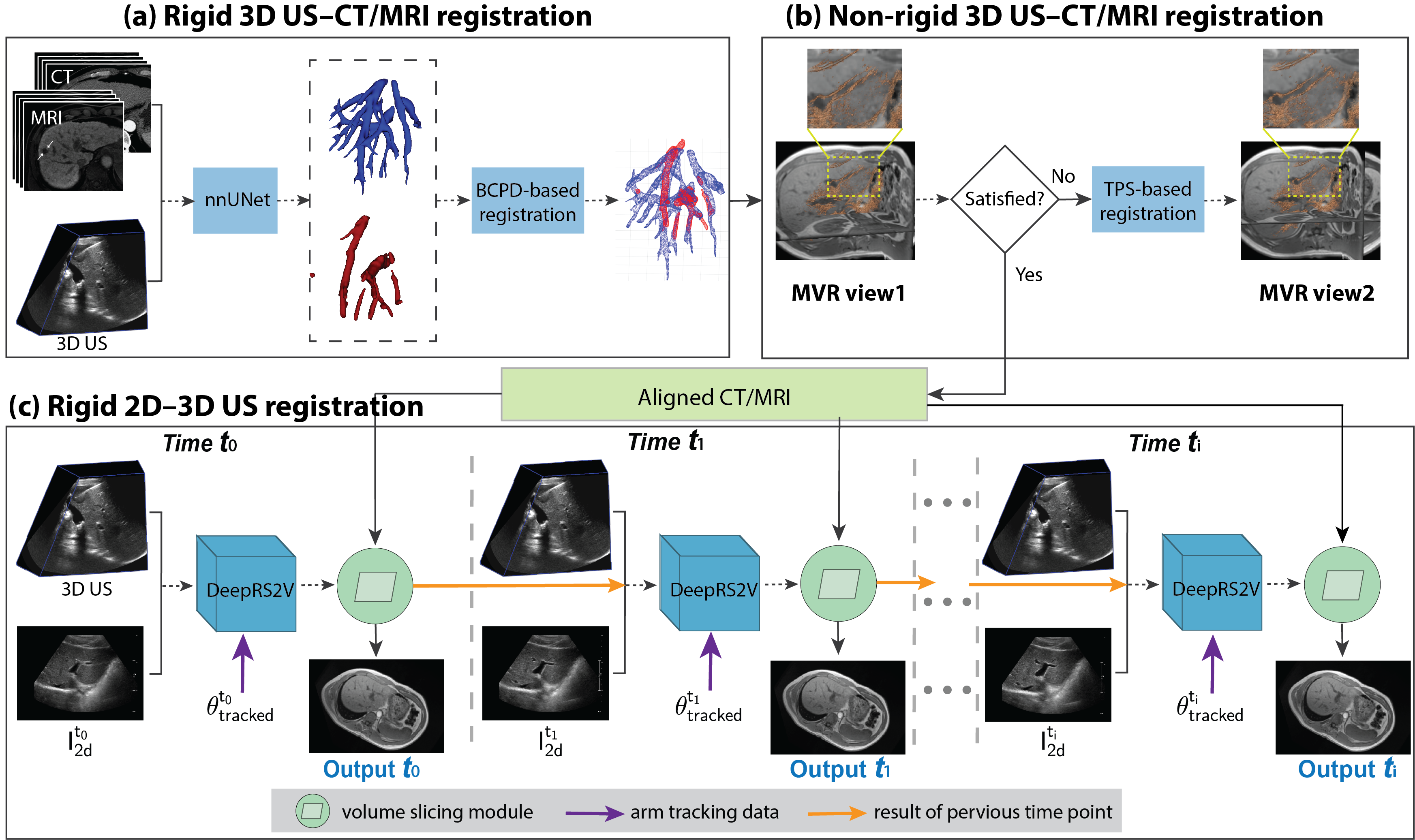}
\caption{Workflow of 2D US–CT/MRI registration.} \label{fig_registration}
\end{figure}

\noindent As shown in Fig.~\ref{fig_registration}, we divided 2D US–CT/MRI registration into three components, as detailed below. 

\subsubsection{Rigid 3D US–CT/MRI}
We first fine-tuned the self-configuring nnUNet algorithm~\cite{isensee2021nnu} for segmenting liver vasculature (\eg, hepatic vessels) from all three modalities. Then, we applied a probabilistic Bayesian coherent point drift (BCPD) approach~\cite{hirose2020bayesian} to rigidly align modality-invariant vessel surface point clouds extracted from CT/MRI and 3D US. \textcolor{black}{Notably}, we constructed a local 3D US liver dataset for segmentation model development, including 95 images collected from 16 healthy volunteers and 13 patients. Further details \textcolor{black}{were described in our previous work}~\cite{xing20233d}.

\subsubsection{Non-rigid 3D US–CT/MRI}
In this step, we introduced Multi-planar and Volume Rendering (MVR), a multimodal image visualization technique that facilitated intuitive alignment assessment. Furthermore, MVR incorporated an interactive non-rigid thin plate spline (TPS)-based registration tool for refining the alignment as needed.

\paragraph{MVR Visualization} As shown in Fig.~\ref{fig_calibration_visualization}b, MVR displays CT/MRI images in three orthogonal planes (\ie, axial, sagittal, and coronal) and renders 3D US using a customized transfer function. The rendered volume (in brown) enhances key liver structures, including hepatic vessels and the liver diaphragm, while excluding blood within the liver vessels to improve vessel alignment detection. To minimize visual obstruction, MVR excludes the rendered volume in front of the viewing plane. For clarity, Fig.~\ref{fig_calibration_visualization}b compares our MVR visualization (b1) with two others (b2 and b3). 
Notably, the rendered content updates automatically according to the viewing plane's position.

\paragraph{TPS-based Registration} If MVR visualization reveals that alignment is clinically unacceptable (see MVR view1 in Fig.~\ref{fig_registration}b), TPS-based interactive deformable registration is applied. 
Initially developed by Bookstein \etal~\cite{bookstein1989principal}, TPS provides an efficient warping approach using a sparse and unstructured set of control points. By incorporating the segmented liver and the physician's workspace, we developed a patient-specific workflow that automatically generates control points to constrain volume deformation. Next, the physician can refine misaligned regions using an intuitive drag-and-place interaction. For effective clinical integration, we also developed a 3D Slicer module, inspired by Gobbi~\etal's work~\cite{gobbi2003generalized}, to perform this registration task. This module is publicly available at GitHub\footnote[1]{https://github.com/Xingorno/MICCAI2025-2DUS-to-CTMRI-Multimodal-Registration.git}.

\subsubsection{Rigid 2D–3D US}
To compensate for internal liver motion due to patient breathing and \textcolor{black}{movement} during the procedure, we developed a time-dependent rigid 2D–3D US image registration workflow. As illustrated in Fig.~\ref{fig_registration}c, at each time point $t_i$, the inputs included a reference 2D US image ($I_{2d}^{t_i}$)  and an intra-procedural 3D US image ($I_{3d}$). By incorporating the 2D US tracking information ($\theta_{tracked}^{t_i}$), we applied deep regression slice-to-volume registration (DeepRS2V) to compute the final transformation. Next, the volume slicing module generated the corresponding CT/MRI multi-planar reformatted (MPR) views. Details of the DeepRS2V and volume slicing module were discussed in our earlier work~\cite{xing2025deep}.


\section{Experiments and Results}
\subsection{System Calibration}



\arrayrulecolor{gray}
\begin{table}[htb]
\caption[Comparison of US image tracking accuracy with/-out US transducer calibration]
{\label{tab_comparison_US_image_tracking_error}
Comparison of US image tracking accuracy with and without US probe calibration. (ED: Euclidean distance, GA: geometric angular error)}
\centering
\resizebox{0.98\textwidth}{!}{
\begin{tabular}{c|cccc|cccc}
\Xhline{1 pt}
\multirow{2}{*}{\makecell{US image tracking}} & \multicolumn{4}{c|}{Translational error (mm)} & \multicolumn{4}{c}{Rotational error ($^\circ$)}\\
{}& {$T_x$} & {$T_y$} & {$T_z$} & {$ED$} & {$R_x$} & {$R_y$} & {$R_z$} & {$GA$}\\
\Xhline{0.6 pt}

\multirow{1}{*}{\makecell{Arm calibration alone}} & \multirow{1}{*}{-3.44} & \multirow{1}{*}{-3.31} & \multirow{1}{*}{-4.01} & \multirow{1}{*}{$7.27 \pm 1.74$} & \multirow{1}{*}{1.74} & \multirow{1}{*}{0.18} & \multirow{1}{*}{0.68} & \multirow{1}{*}{$3.46 \pm 1.39$} \\
\hline

\multirow{1}{*}{\makecell{Arm \& Probe calibration}} & \multirow{1}{*}{-0.14} & \multirow{1}{*}{-0.07} & \multirow{1}{*}{-0.73} & \multirow{1}{*}{$\mathbf{2.79 \pm 1.53}$} & \multirow{1}{*}{0.94} & \multirow{1}{*}{-0.60} & \multirow{1}{*}{-0.10} & \multirow{1}{*}{$\mathbf{2.92 \pm 1.23}$} \\
\Xhline{1 pt}
\end{tabular}}
\end{table}

\noindent Fig.~\ref{fig_calibration_visualization}a shows the experimental setup for calibration, which was verified by measuring the tracking pose error, defined as translation and rotation differences between the tracked and ground-truth US image planes. The ground-truth plane is defined by 3D landmarks, $p_{3d}^i$, derived from multiple intersection points (see $p^1,p^2,…,p^5$ in Fig.~\ref{fig_calibration_visualization}a) via triangular similarity. The tracked image plane is determined after system calibration. 
As shown in Tab.~\ref{tab_comparison_US_image_tracking_error}, US image tracking with probe calibration yielded a Euclidean translational error of \qty{2.79}{\milli\meter} $\pm$ \qty{1.53}{\milli\meter}, approximately one third of the arm calibration alone. Since registration is not involved in this evaluation (error = 0), this result represents the \textcolor{black}{theoretical} lower bound of CT/MRI MPR accuracy (calibration + registration).

\subsection{2D US–CT/MRI Registration}
\begin{figure}
\centering
\includegraphics[width=0.65\textwidth]{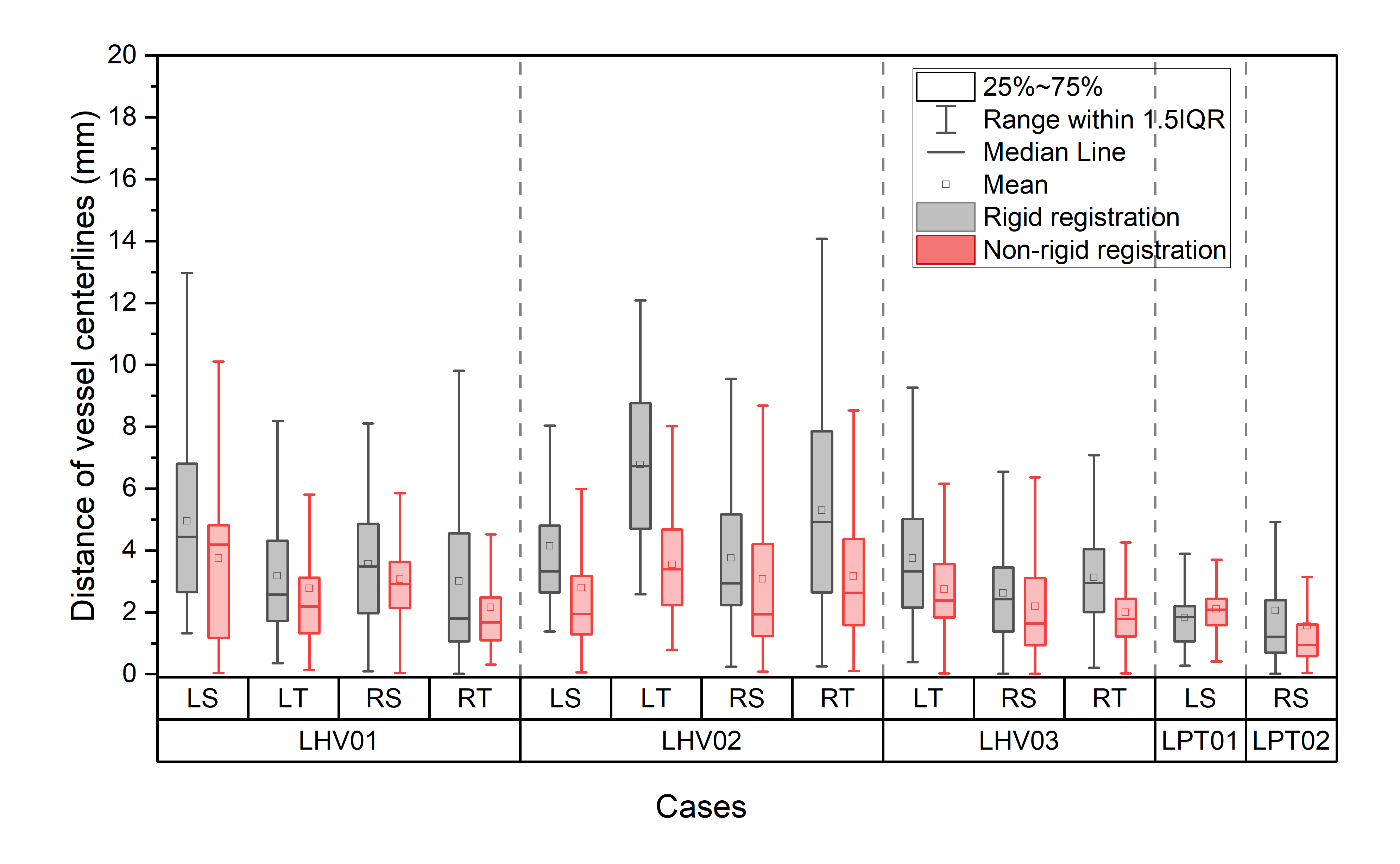}
\caption{Comparison of rigid and non-rigid registration on healthy volunteers and patients. (LHV: liver healthy volunteer, LPT: liver patient)} \label{fig_comparion_rigid_deformable}
\end{figure}

\subsubsection{3D US–CT/MRI}
Since tumours were absent in healthy volunteers, vessel centreline distance ($D_{cl}$) served as a surrogate metric for registration accuracy. We evaluated 13 cases from 3 healthy volunteers and 2 patients. 3D US images were acquired from four liver regions with minimal overlap between images, including right-transverse (RT), right-sagittal (RS), left-transverse (LT), and left-sagittal (LS). Due to radiation exposure from CT, Research Ethics Board allowed only the use of MRI for the healthy volunteer study. 
For patients LPT01 and LPT02, MRI and CT images were acquired, respectively. In Fig.~\ref{fig_comparion_rigid_deformable}, all cases except LPT01-LS showed a reduction in the mean $D_{cl}$ after non-rigid registration, ranging from \qty{1.58}{\milli\meter} to \qty{3.75}{\milli\meter}, corresponding to a \SIrange{13.0}{47.7}{\percent} reduction compared to \textcolor{black}{the} rigid registration. Notably, variability in registration error across cases \textcolor{black}{persisted} after non-rigid correction. Moreover, rigid registration in LPT01-LS achieved good alignment (\qty{1.83}{\milli\meter} $\pm$ \qty{0.97}{\milli\meter}), thereby rendering deformable correction less perceptible. The rigid registration step (Fig.~\ref{fig_registration}a) took < \qty{1}{\minute} and the interactive non-rigid step (Fig.~\ref{fig_registration}b) requires physicians' intervention, with a duration ranging from seconds to minutes. 

\subsubsection{2D US–3D US}
This step was previously reported in our earlier work~\cite{xing2025deep}. For reference, we re-visit the registration result, which achieved a mean Euclidean distance (ED) error of \qty{2.28}{\milli\meter} $\pm$ \qty{1.81}{\milli\meter} and a mean geometric angular (GA) error of \qty{2.99}{\degree} $\pm$ \qty{1.95}{\degree}. The runtime per pair was approximately \qty{0.22}{\second}.

\subsection{System Workflow Validation}

\arrayrulecolor{gray}

\begin{table}[htb]
\caption[Results of CT/MRI planar reformatting on five cases from four healthy volunteers. ]
{\label{tab_healthy_volunteer_mpr_results}
Results of MRI MPR on five cases from four healthy volunteers. (\textbf{Regi-a}: Rigid 3D US–MRI; \textbf{Regi-b}: Non-rigid 3D US–MRI; \textbf{Regi-c}: Rigid 2D–3D US)}
\centering
\resizebox{0.98\textwidth}{!}{
\begin{tabular}{c|c|c|c|c|c|c}
\Xhline{1 pt}
\multicolumn{2}{c|}{\textbf{Integration Components}} & {LHV-I-1} & {LHV-I-2} & {LHV-II} & {LHV-III} & {LHV-IV}\\
\Xhline{0.6 pt}

\multirow{1}{*}{\centering \makecell{\textbf{Regi-a}}} & {\makecell{$D_{cl}$ (mm)}} & \multicolumn{2}{c|}{$3.73 \pm 2.69$} & {$3.48 \pm 4.15$} & {$2.74 \pm 2.00$} & {$3.23 \pm 2.53$}\\
\hline

\multirow{2}{*}{\centering \makecell{\\ \textbf{Regi-b}}} & {\makecell{$D_{cl}$ (mm)}} & \multicolumn{2}{c|}{$2.34 \pm 1.57$} & {$2.03 \pm 2.75$} & {$1.75 \pm 1.27$} & {$1.81 \pm 1.68$} \\ \cline{2-7}
{} & {\makecell{Error reduced}} & \multicolumn{2}{c|}{\textdownarrow \qty{37.3}{\percent}} & {\textdownarrow \qty{41.7}{\percent}} & {\textdownarrow \qty{36.1}{\percent}} & {\textdownarrow \qty{44.0}{\percent}} \\
\hline

\multirow{3}{*}{\centering \makecell{\textbf{Regi-c}}} & {\makecell{No. of US frames}} & {68} & {35} & {41} & {18} & {31} \\ \cline{2-7}
{}& {\makecell{No. of landmarks}} & {6} & {5} & {9} & {5} & {6} \\ \cline{2-7}
{}& {TRE (mm)} & {$1.37 \pm 0.80$} & {$1.40 \pm 1.86$} & {$1.36 \pm 1.41$} & {$3.41 \pm 4.76$} & {$0.89 \pm 0.76$} \\ 
\hline

\multirow{3}{*}{\centering \makecell{ \textbf{MPR} \\ \textbf{views}}} & {\makecell{No. of landmarks}} & {6} & {4} & {9} & {4} & {6} \\ \cline{2-7}
{} & {\makecell{Rigid: LD (mm)}} & {$3.52 \pm 1.11$} & {$3.52\pm2.08$} & {$6.82 \pm 4.00$} & {$4.57 \pm 3.86$} & {$2.73 \pm 1.91$}\\ \cline{2-7}
{}& {\makecell{Non-rigid: LD (mm)}} & {$\mathbf{3.15 \pm 1.17}$} & {$\mathbf{3.49 \pm 1.72}$} & {$\mathbf{4.53 \pm 3.35}$} & {$\mathbf{3.88 \pm 2.96}$} & {$\mathbf{2.35 \pm 0.89}$} \\
\Xhline{1 pt}
\end{tabular}}
\end{table}

\begin{figure}[htbp]
\centering
\includegraphics[width=0.98\textwidth]{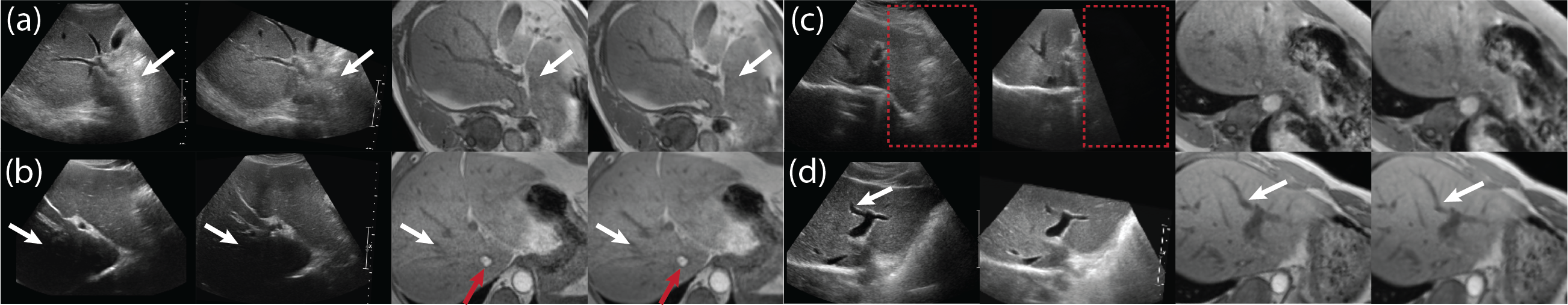}
\caption{Results of US–CT/MRI registration. (For each case, from left to right: original US, 3D US MPR, rigid and non-rigid CT/MRI MPR.)} \label{fig_registration_results}
\end{figure}

\noindent Four additional healthy volunteers (LHV-I to -IV) were recruited to validate the entire workflow. For each subject, we acquired an MRI image, a 3D US image, and a US video. 
Tab.~\ref{tab_healthy_volunteer_mpr_results} shows that non-rigid registration resulted in a mean $D_{cl}$ of $\sim$2 mm across all five cases, corresponding to a $\sim$40\% reduction in error compared to rigid registration. For 2D–3D US registration, liver vessel bifurcation points were manually selected as landmarks to compute the target registration error (TRE). The same evaluation was applied to subsequent CT/MRI MPR, but the evaluation metric was referred to as landmark distance (LD) instead of TRE. Tab.~\ref{tab_healthy_volunteer_mpr_results} shows that all video clips except LHV-III had TRE < \qty{1.5}{\milli\meter}. In addition, CT/MRI MPR (non-rigid) achieved a mean LD < \qty{4.0}{\milli\meter}, which is deemed clinically acceptable~\cite{laimer2020minimal}. In Fig.~\ref{fig_registration_results}a and b, regions affected by US shadowing artifacts (white arrow) were restored in the registered MRI images. Complementary liver structures, such as the inferior vena cava (red arrow), could aid procedural planning for guidance safety. Our 2D–3D US registration also achieved robust alignment in cases with limited overlap (Fig.~\ref{fig_registration_results}c) and deformation (Fig.~\ref{fig_registration_results}d).

\section{Discussion}

\noindent In clinical practice, a 2D US–CT/MRI registration error of less than \qty{5}{\milli\meter} is considered an acceptable threshold~\cite{laimer2020minimal}, given the widely used 5–10 \si{\milli\meter} safety margin for complete tumour eradication~\cite{shady2018percutaneous}. Results from healthy volunteers confirmed that our approach meets this requirement. Additionally, the MVR visualization assists interactive TPS-based deformable correction when needed. Given its effectiveness in handling challenging cases, this step adds minimal time to the workflow beyond the necessary alignment check in a workflow without this non-rigid step. The registration workflow for applicator guidance ran at \qty{0.22}{\second} per image pair, enabling dynamic observation of the aligned CT/MRI. Although this runtime is clinically acceptable, future work will include patient studies to assess whether \textcolor{black}{further} runtime optimization is needed.

\section{Conclusion}
This study proposed a novel framework for 3D US integration in percutaneous liver tumour ablation, aiming to improve tumour identification and multimodal image visualization. The results demonstrated the efficacy and clinical applicability of the proposed 2D US–CT/MRI registration workflow in enhancing tumour localisation. Additionally, this work highlighted the potential of the 3D US integration to facilitate and improve percutaneous tumour ablation. 

%
%

    

\begin{credits}
\subsubsection{\ackname} This study is funded by Natural Sciences and Engineering Research Council
of Canada (1248179);
Ontario Research Fund-Research Excellence;
Canadian Foundation for Innovation (36199); Ontario Institute for Cancer
Research (RA262); Canadian
Institute of Health Research (154314, 143232 and 201409).

\subsubsection{\discintname}
The authors have no competing interests to declare that are
relevant to the content of this article.

\subsubsection{Ethical Approval.}
Patient and volunteer trials were approved by the Western University Health Science Research Ethics Board and conducted in accordance with the TCPS 2 guidelines.
\end{credits}

%
%
%
\bibliographystyle{splncs04}
\bibliography{mybibliography}
%





\end{document}